\documentclass[conference]{IEEEtran}
\IEEEoverridecommandlockouts
\usepackage{cite}
\usepackage{amsmath,amssymb,amsfonts}
\usepackage{algorithmic}
\usepackage{graphicx}
\usepackage{textcomp}
\usepackage{xcolor}
\usepackage{subcaption}
\usepackage[backref]{hyperref}
\def\BibTeX{{\rm B\kern-.05em{\sc i\kern-.025em b}\kern-.08em
    T\kern-.1667em\lower.7ex\hbox{E}\kern-.125emX}}
\begin{document}

\title{{Early ChatGPT User Portrait through the Lens of Data}\\
}

\author{\IEEEauthorblockN{1\textsuperscript{st} Yuyang Deng}
\IEEEauthorblockA{\textit{Dept. of Applied Mathematics and Statistics} \\
\textit{Johns Hopkins University}\\
Baltimore, USA \\
ydeng36@jh.edu}
\and
\IEEEauthorblockN{2\textsuperscript{nd} Ni Zhao}
\IEEEauthorblockA{\textit{Dept. of Biostatistics} \\
\textit{Johns Hopkins University}\\
Baltimore, USA \\
nzhao10@jhu.edu}
\and
\IEEEauthorblockN{3\textsuperscript{rd} Xin Huang}
\IEEEauthorblockA{\textit{Dept. of Computer and Information Sciences} \\
\textit{Towson University}\\
Baltimore, USA \\
xhuang@towson.edu}
\and
}

\IEEEoverridecommandlockouts
\IEEEpubid{\makebox[\columnwidth]{979-8-3503-2445-7/23/\$31.00~\copyright2023 IEEE \hfill} \hspace{\columnsep}\makebox[\columnwidth]{ }}

\maketitle

\begin{abstract}
\\
Since its launch, ChatGPT has achieved remarkable success as a versatile conversational AI platform, drawing millions of users worldwide and garnering widespread recognition across academic, industrial, and general communities. This paper aims to point a portrait of early GPT users and understand how they evolved. Specific questions include their topics of interest and their potential careers; and how this changes over time. We conduct a detailed analysis of real-world ChatGPT datasets with multi-turn conversations between users and ChatGPT. Through a multi-pronged approach, we quantify conversation dynamics by examining the number of turns, then gauge sentiment to understand user sentiment variations, and finally employ Latent Dirichlet Allocation (LDA) to discern overarching topics within the conversation. By understanding shifts in user demographics and interests, we aim to shed light on the changing nature of human-AI interaction and anticipate future trends in user engagement with language models.




\end{abstract}

\begin{IEEEkeywords}
ChatGPT, Natural Language Processing, Sentiment Analysis, Latent Dirichlet allocation, Topic Modeling
\end{IEEEkeywords}

\section{Introduction}

 ChatGPT is an advanced AI language model developed by OpenAI, utilizing the revolutionary transformer architecture and self-attention mechanisms for exceptional language understanding and generation. It leverages large-scale language modeling, fine-tuned with techniques like Reinforcement Learning from Human Feedback (RLHF), to deliver nuanced and contextually aware responses across a wide array of conversational topics. Using multiple layers of self-attention and feed-forward neural networks, ChatGPT is able to capture the context and relationships between words in the input sequence, and in some sense, truly ``understand'' the user inquiry. This profound ``understanding'' is a result of comprehensive training across a diverse array of data sources, encompassing the vast expanse of the internet, articles, and books. ChatGPT's model is pre-trained, affording it the ability to promptly and almost instantaneously provide feedback.  On the other hand,  ChatGPT is also limited: ChatGPT lacks real-time internet access for up-to-the-minute information, and its training data only extends up to September 2021. 

So far, ChatGPT has gained remarkable success in amassing users. According to Reuter's report\cite{Reuters}, ChatGPT achieved a milestone of 1 million users in just five days after its initial release in November 2022 and gained 100 million active users by January 2023. By comparison, it took Instagram approximately 2.5 months to reach 1 million downloads, whereas Netflix had to wait around 3.5 years to reach 1 million users. Importantly, the user base has continued to expand steadily. As of the latest available data up to August 2023, ChatGPT boasts a user base exceeding 180.5 million.  This substantial repository of user behavior data provides an invaluable opportunity to glean insights into user characteristics, preferences, and areas of interest. Such analyses hold great potential for the benefit of developers, researchers, educators, and businesses, offering a clearer understanding of AI-chatbot interactions and pinpointing avenues for growth or areas of concern.

 In this paper, we provide an in-depth analysis of the ChatGPT conversations that are collected by shareGPT~\cite{sharegpt}, an open-source Chrome Extension that allows users to share their ChatGPT conversation with one click. With this extension, ChatGPT users agree to contribute their conversations to a public database. Two datasets from shareGPT were analyzed, with the first one (shareGPT52K) collected before April 2023 and the second one (shareGPT92K) collected from its launch to June 2023.

Our primary contribution is the comprehensive analysis of early ChatGPT user interactions to understand the conversation dynamics and user portrait. In Section III, we present the key statistical attributes of the shareGPT dataset. In Section IV, we gain insights into the sentiment aspects of the conversations by exploring how users' sentiments evolve during interactions with ChatGPT. In Section V, we utilize the Latent Dirichlet Allocation (LDA) technique to investigate the specific topics that users express interest in when engaging with ChatGPT.

\section{Related Work}

Although ChatGPT is a relatively recent development, it has opened new avenues in natural language processing (NLP) and artificial intelligence (AI) research. Research focusing on ChatGPT has largely revolved around three main areas: exploring its potential applications in specific domains, evaluating its effectiveness in a variety of conversational scenarios, and assessing the potential challenges and risks associated with the use of ChatGPT.

In the first theme, researchers have assessed the use of ChatGPT in various fields, including medical care(\cite{rao2023evaluating}, \cite{gabrielson2023harnessing}, \cite{hulman2023chatgpt}), education(\cite{rudolph2023chatgpt},\cite{frieder2023mathematical}\cite{pardos2023learning}), reasoning\cite{bang2023multitask}, software development\cite{Jalil_2023}, translation\cite{jiao2023chatgpt}, and scientific research(\cite{stokel2023chatgpt}, \cite{korinek2023language}), and highlighted the feasibility of using ChatGPT in these areas. 
The second theme delves into the strengths and weaknesses of ChatGPT.  Guo \cite{guo2023close}  conducted a comparative analysis between ChatGPT-generated responses and human expert-generated answers for various questions, concluding that ChatGPT provides better-organized, more detailed, and less biased responses, despite occasional fabrication of facts in its answers. Borji\cite{borji2023categorical}  demonstrated ChatGPT's suboptimal performance in responding to questions in eleven distinct categories, including reasoning, factual accuracy, mathematics, coding, and bias, cautioning that ChatGPT may struggle to generate creative solutions for novel problems. In the third theme, scholars have focused on ethical concerns surrounding ChatGPT, including its potential misuse in law school exams \cite{choi2023chatgpt}, plagiarism detection in education \cite{khalil2023will}, and the potential for misuse in medical practices \cite{shen2023chatgpt}.



Despite these notable advancements, a significant gap in the research landscape persists. There has been a lack of studies that systematically evaluate the user profiles of ChatGPT, exploring users' specific interests and potential shifts in those interests over recent months. Consequently, this study represents the first of its kind, aimed specifically at bridging this gap in our understanding of ChatGPT's user behavior.


\begin{table}[htbp]
\caption{Descriptive statistics of conversation lengths for the shareGPT92K Dataset} 
\label{tab:description}
\begin{center}
\begin{tabular}{lrl}
\hline 
\textbf{Parameter}& \textbf{Value} \\
\hline
\# conversations& 87,957\\
Total \# turns & 1,449,998\\
Average \# turns per conversation & 16.48 \\
Median \# turns per conversation & 6.00 \\
Maximum \# turns per conversation & 998 \\
\# of conversations that finish within two turns & 24366 (27.7\%) \\
\# of conversations that finish within four turns & 37684 (42.8\%) \\
\# of conversations that have even turns & 84915 (96.5\%) \\
\hline
\end{tabular}
\end{center}
\end{table}

\section{Conversation Feature Analysis}
\subsection{Dataset Description}
Two ChatGPT user interaction datasets, shareGPT52K and shareGPT92K collected by shareGPT, were downloaded from the Hugging Face data repository~\cite{huggingface}. These datasets were compiled using shareGPT~\cite{sharegpt}, an open-source Chrome Extension designed to facilitate the convenient sharing of ChatGPT conversations. In essence, users willingly consented to the collection and sharing of their interactions by employing shareGPT. This included the recording and downloading of their user IDs and the entirety of their interactions with ChatGPT. The two datasets were retrieved in April and June of 2023, respectively. It is worth noting that the shareGPT52K is a subset of the shareGPT92K dataset. 

Within this study, our primary focus centers on the analysis of the shareGPT92K dataset due to its provision of more comprehensive and extensive data than the shareGPT52K dataset. However, for the purpose of delineating user dynamics (as discussed in Section V.C) and conducting topic analysis, we will analyze the characteristics of these two datasets respectively in addition to the joint analysis.

\begin{figure}[htbp]
\centerline{\includegraphics[scale=0.4]{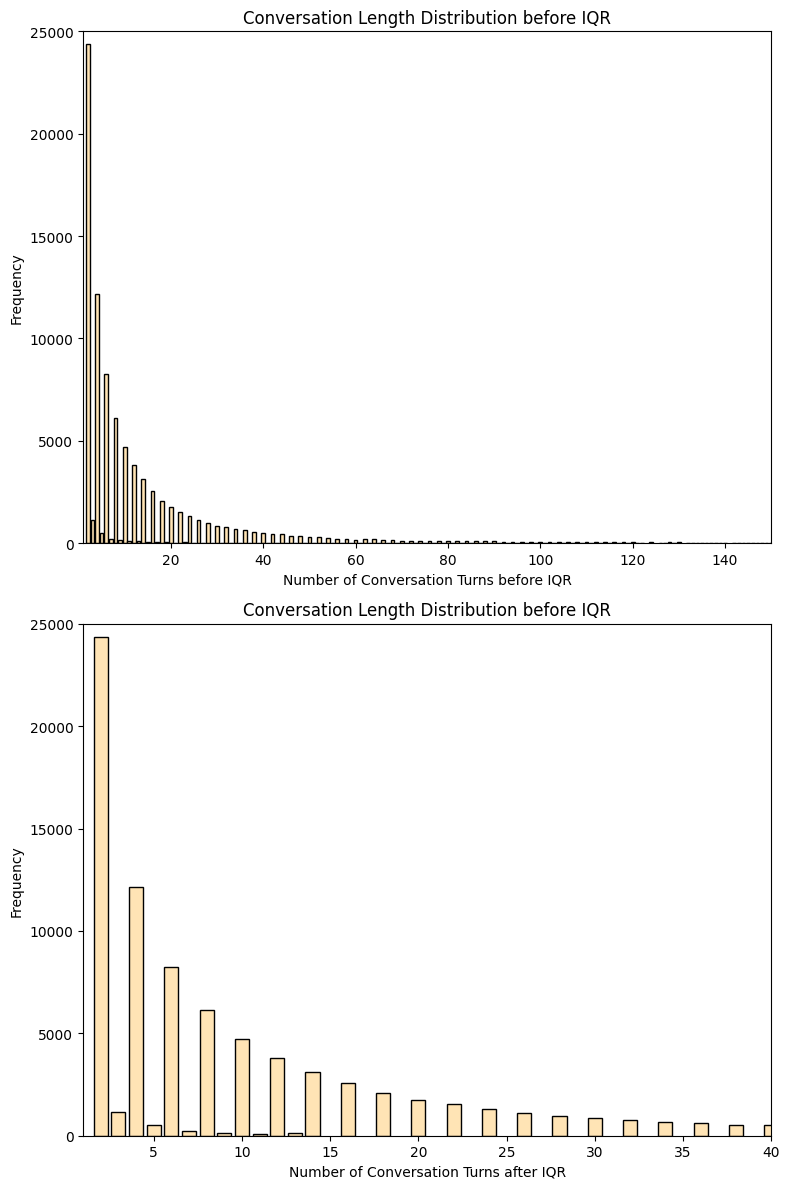}}
\caption{Conversation length distribution in the shareGPT92K dataset, before(top panel) and after(bottom panel) IQR (Interquartile Range) filtering. For better visualization, we do not show the distribution of lengths greater than 140 with a frequency of only 1.}
\label{fig:lengthDistribution}
\end{figure}





\subsection{Length Distribution}

We omit 2,708 invalid conversations, which are defined as those instances where conversations consist of only a single turn initiated by a human with no response from ChatGPT, likely due to connection issues, the shareGPT92K dataset comprises 87,957 valid conversations for our analysis. The table \ref{tab:description} and the top panel in Fig~\ref{fig:lengthDistribution} provide a summary of the length characteristics of these conversations. Evidently, the majority of these conversations exhibit an even number of turns, implying that ChatGPT consistently responds to inquiries or conversations initiated by humans. Specifically, 27.7\% of these conversations concluded within two turns in which ChatGPT successfully addressed the user's question within just one iteration. Additionally, 42.8\% of these interactions conclude within four turns, suggesting ChatGPT's ability to promptly address users' inquiries. Over 65.5\% of conversations finish within 10 turns.




In the dataset, some conversations have more than 700 turns, suggesting that the users didn't initiate a ``new chat'' after a topic was finished. Considering that these conversations may skew the results of our analysis, we excluded them during the visualization process. To do so, we employed the Interquartile Range (IQR) method. Specifically, we consider conversations whose lengths (number of turns) are 1.5 times IQR below the first quartile or above the third quartile as outliers and removed these conversations. In total, we removed 8412 such abnormal records. The bottom panel of Fig~\ref{fig:lengthDistribution} showed the length distribution after this filtering, which showed a clearer view of the aforementioned trend.





\subsection{Conversation Languages Distribution}

\begin{figure}[htbp]
\centerline{\includegraphics[scale=0.4]{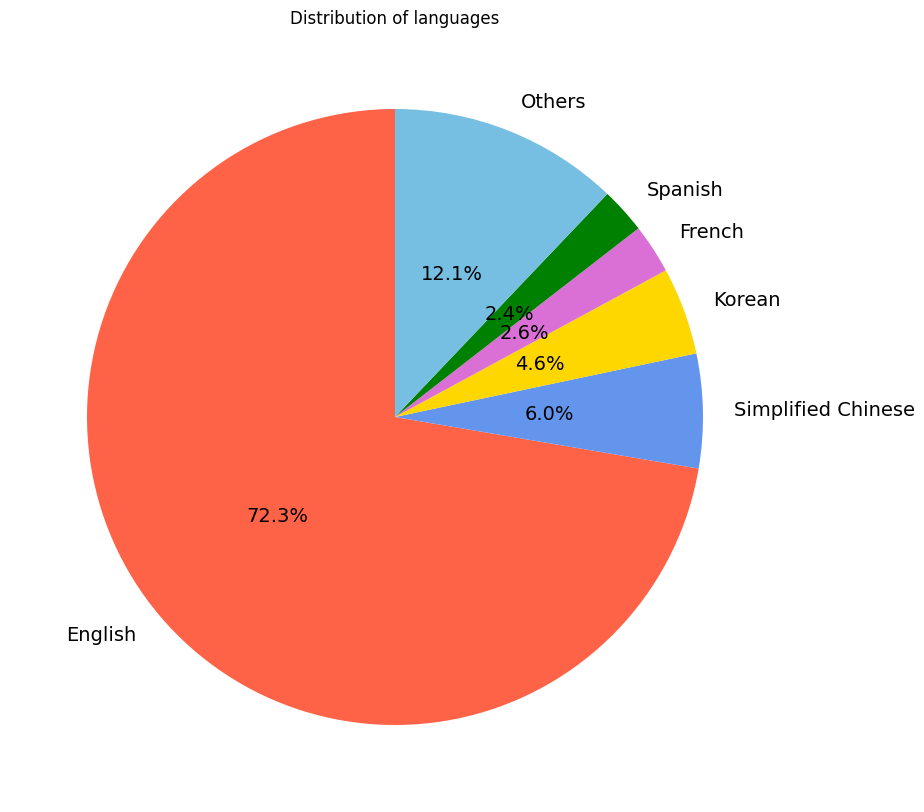}}
\caption{Distribution of Conversation Language of 92K Dataset.}
\label{fig:languagePie}
\end{figure}

As a globally oriented product, ChatGPT has accumulated conversations in many languages. We employ a public language detection tool in Java~\cite{language-detection} to ascertain the languages used in the shareGPT database. In total, we have identified 45 languages, with the most prominent ones being English, Simplified Chinese, Korean, French, and Spanish (as illustrated in Fig~\ref{fig:languagePie}). Remarkably, English accounts for over 70\% of all the conversations. Approximately 6\% of the conversations are conducted in Simplified Chinese, making it the second most prevalent language in ChatGPT interactions, despite ChatGPT not being directly available in mainland China. Several factors could contribute to this phenomenon. These conversations might originate from users in regions where Simplified Chinese is commonly spoken, such as Singapore, Malaysia, or overseas Chinese communities. It's also plausible that some users from mainland China have found ways to bypass restrictions, such as through the use of virtual private networks (VPNs) or other tools. Furthermore, the Others language category, encompassing 40 languages beyond the top five, constitutes 12.1\% of all conversations. Nevertheless, the prevalence of these diverse languages within ChatGPT underscores its multilingual capabilities and the global reach of AI language models, which are capable of serving users from various linguistic backgrounds and regions.





\section{Sentiment Analysis}


Sentiment analysis plays a crucial role in understanding ChatGPT user conversations. By analyzing the sentiment in user interactions, developers can gain insights into how users feel about their interactions with ChatGPT. Positive sentiments might indicate satisfaction and engagement, while negative sentiments can highlight areas needing improvement.

Sentiment analysis is instrumental in deciphering the "emotions" underlying ChatGPT user interactions. Given its role as a large language model, it's essential to understand the ability of ChatGPT to possess the capability to comprehend and interpret human emotional nuances. 

\begin{figure*}[htbp]
\centerline{\includegraphics[scale=0.5]{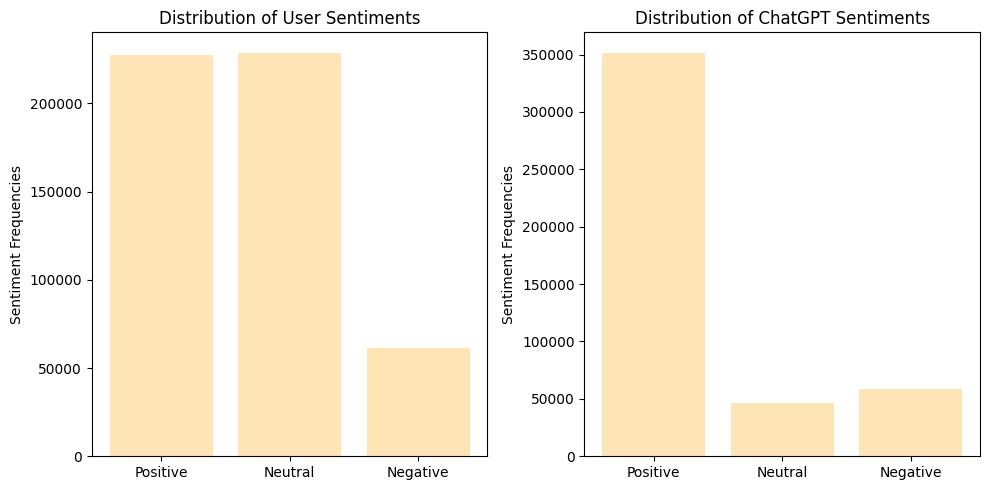}}
\caption{Sentiment Analysis of ChatGPT user interactions. The left and right panels show the sentiment distributions of user inquiry and ChatGPT response, respectively.}
\label{fig3}
\end{figure*}

\begin{figure*}[htbp]
\centerline{\includegraphics[scale=0.37]{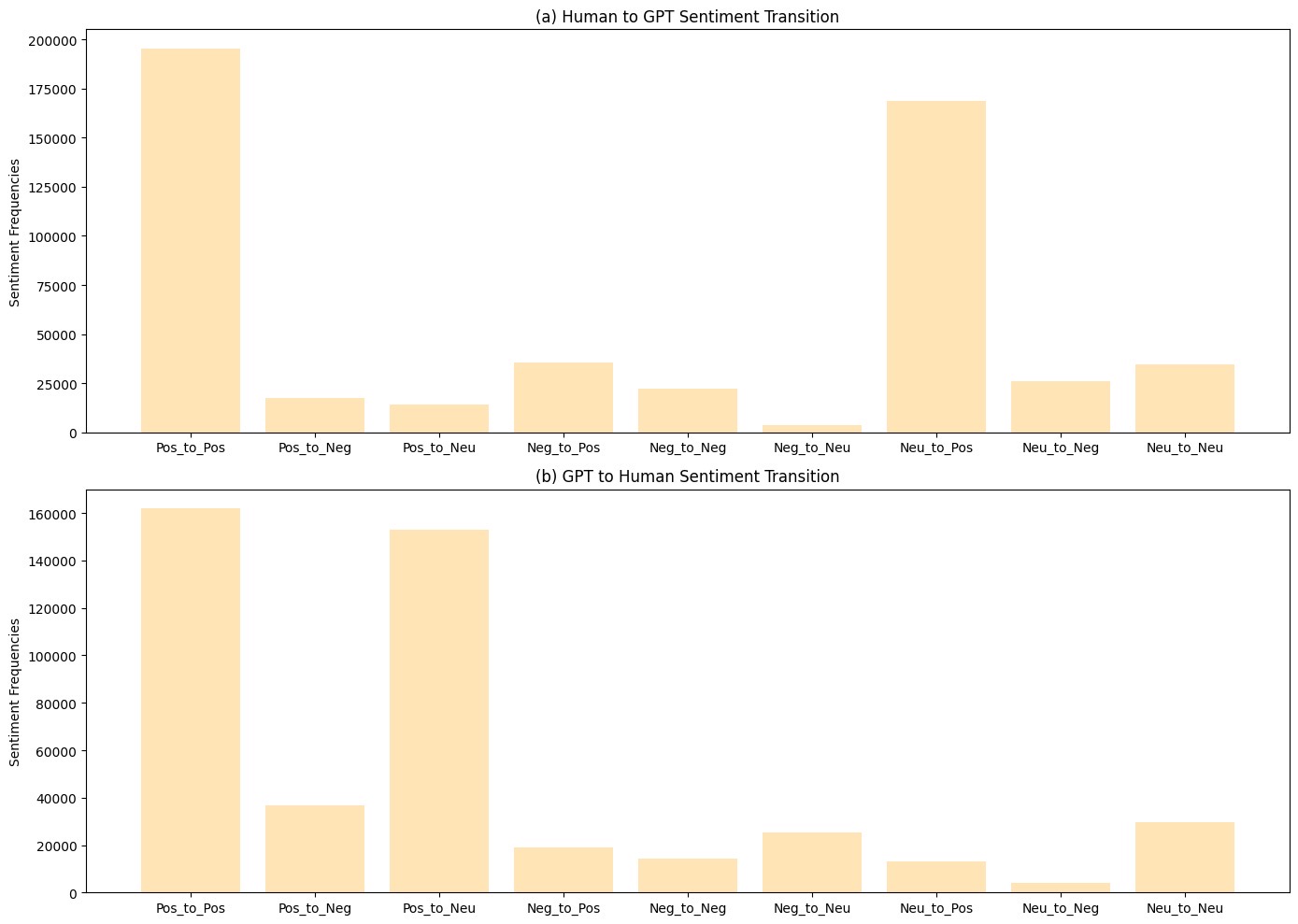}}
\caption{Sentiment Analysis of ChatGPT-Users interactions. The top panel (a) shows the sentiment transition from human question to ChatGPT answer. The bottom panel (b) shows the sentiment transition from the ChatGPT answer to the human's follow-up question. For better visualization, we use `Pos' to represent `Positive', `Neg' to represent `Negative', and `Neu' to represent `Neutral'.}
\label{fig4}
\end{figure*}

As emotional creatures, humans infuse their inquiries with feelings, and ChatGPT, being trained on the extensive human-generated text and fine-tuned through human feedback, reflects this emotional aspect. Consequently, our interest lies in exploring the emotional insights of these interactions, with a specific focus on analyzing the sentiment distributions and sentiment shifts.


\begin{table*}[htbp]
\caption{Conversation Topics in 52K, 42K, and 92K Datasets}
\begin{center}
\begin{tabular}{c c c c}
\hline
\textbf{Topics}& \textbf{52K} & \textbf{40K} & \textbf{92K}\\
\hline
1& code,  file, return, error, function & file, data, code, import, value & file, return, import, code, error\\
 & data, import, using, line, create & return, error, model, using, text & value, line, function, data, class\\
2& system, service, company,data, model & URL, date, result, write, search & model, write, data, using, user\\
 & business, team, provide, information, work & current, English, using, query, web & example, contract, information, description, answer\\
3& people, time, make, think, way & make, time, get, way, people & date, URL, result, search, using\\
 & get, day, know, back, story & know, think, world, back ,thing & subject, query, web, provided, current\\
4& country,  state,  contract, work, information& URL, result, search, query, write & business, company, service, product, value\\
 & Russian, answer, evidence, petition, Ukrainian & web, messi, Korean, e, provided & credit, team, balance, debit, customer\\
5& write, continue, make, prompt, search & date, contract, write, information, topic & make, time, write, people, way\\
 & content, answer, give, word, result & description, credit, system, post, balance & get, know, think, could, first\\
\hline
\end{tabular}
\label{tab1}
\end{center}
\end{table*}

Given that ChatGPT can notably adjust its responses based on human directives, and sometimes, even apologies to users or confusing right and wrong, our research first dived into the prevalent sentiment states (Positive, Neutral, Negative) of users and ChatGPT, then we examined whether ChatGPT(or human) is likely to modify the sentiment states of its responses based on the sentiments detected in previous user(ChatGPT) questions(answers).

In our experiment, we applied the sentiment analysis module in NLTK\cite{nltksentiment} for analysis and got the below result.

From the left and right panels of Fig~\ref{fig3}, it is observed that the emotional tone of questions posed by users predominantly aligns with positive and neutral sentiments, indicating a minimal inclination among users to express negative emotions in their inquiries. In response, ChatGPT predominantly generates answers with a positive sentiment, suggesting its predisposition towards producing emotionally positive content. 

Moreover, We are interested in what kind of responses ChatGPT will give based on the sentiments in the user's questions, as well as the inverse procedure, which led us to detect the sentiment fluctuations within conversations. By comparing the top panel (a) and bottom panel (b) in Fig~\ref{fig4}, we found that although positive-to-positive question-answer(answer-question) pairs occupy the main position with a ratio of 37.7\% and 35.4\% respectively, the main difference is that ChatGPT is likely to give positive answers when given a neutral question (32.6\%), while people may turn back to a neutral tone in the follow-up questions when given a positive answer (33.5\%). This interesting result is consistent with our previous observations in Fig~\ref{fig3} that ChatGPT has a trend to give a positive response.



\section{Topics Modeling and Visualization}
After conducting basic research on the inherent characteristics of our dataset, in this section, we delved into the distribution of topics presented in human-machine dialogues. We clean the data by removing noises and unnecessary information, using the word cloud to find the most frequent words, and implementing Latent Dirichlet Allocation (LDA) to separate datasets into several topics.

\subsection{Preprocessing}
Before implementing advanced analysis, we first clean the data, it includes processes of
\begin{itemize}
    \item Text Cleaning: removing noise(URL, HTML tags, and number expressions) and lowercasing.
    \item Tokenization: breaking down text into individual words or phrases.
    \item Stopword Removal: filtering out common words, like 'and', 'the', and 'is', which are often insignificant for our analysis.
    \item Stemming: reducing words to their base or root form, in order to simplify our text analysis task.
    \item Lemmatization: converting words to their base or dictionary form.
\end{itemize}

\subsection{WordCloud}
\begin{figure}[htbp]
\centerline{\includegraphics[scale=0.25]{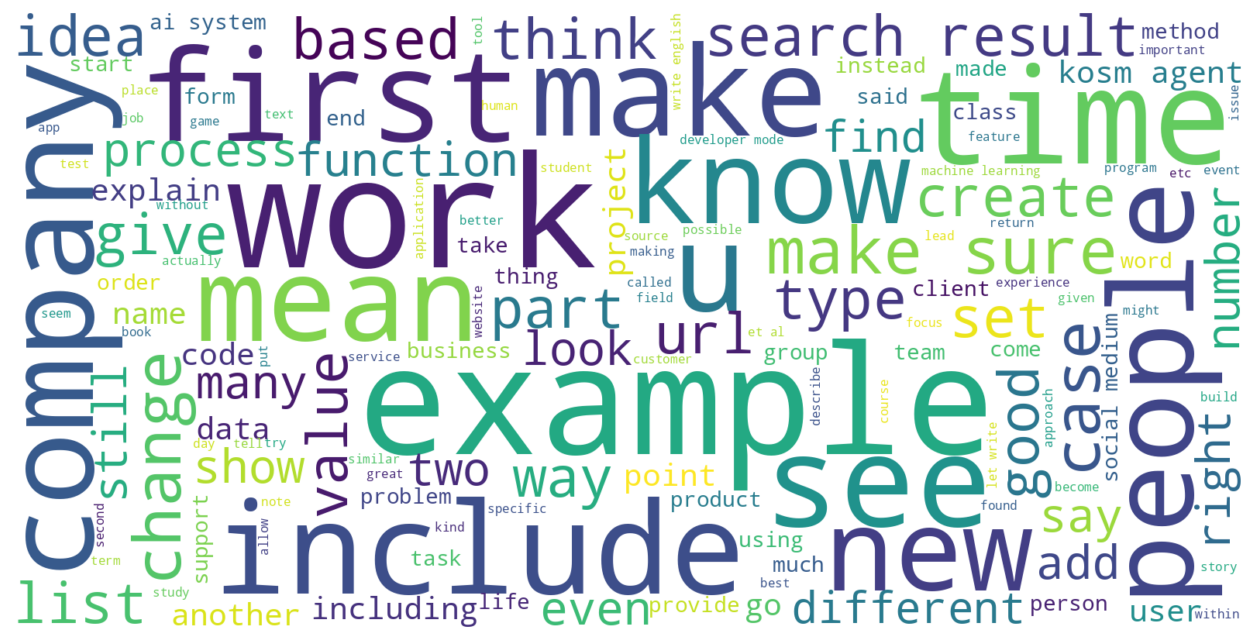}}
\caption{WordCloud of Human Questions.}
\label{fig5}
\end{figure}

A word cloud is a visual representation of text data where the frequency of each word is depicted by its size and prominence. It provides a quick glance at the most mentioned terms, highlighting the main themes or patterns in the data.
In this part, we apply word cloud to our dataset to find out the most frequently used words by humans. As Fig\ref{fig5} shows, when asking ChatGPT questions, humans are likely to require ChatGPT to give them an `example', `case' or `new idea' about their `company' or `work', with clear instruction using words like `first', `make', `new' or `include'. This result is also consistent with the long-standing habit of humans asking questions to other people or asking questions on search engines/forums.

\subsection{Topic Distribution}
Latent Dirichlet Allocation (LDA)\cite{blei2003latent} is a generative probabilistic model used for topic classification in text documents. It assumes documents are mixtures of topics and represents each word as belonging to a particular topic. By analyzing word frequencies, LDA discerns the mixture of topics within a document, effectively uncovering hidden thematic structures in large text collections. We use LDA to implement unsupervised learning on our dataset and choose coherence scores as evaluation metrics. After comprehensively considering the model performance and visualization intuitiveness, we separated the dataset into 5 topics.

As illustrated in Table\ref{tab1}, when separated into 5 topics, we find users' questions mainly lies on programming(Topics 1 in 3 dataset), finance(Topics 2, 5, 4 in 52K, 40K, and 92K dataset respectively), international news(Topics 4 in 52K, 40K dataset respectively), interpersonal communication or writing(Topics 3 in 52K, 40K dataset, topics 5 in 92K dataset), etc.

The predominant theme across 3 datasets centers on coding and data processing, aligning with the widespread astonishment and subsequent adoption of ChatGPT for its remarkable coding proficiency. After its release, a significant number of users turned to ChatGPT, impressed by its ability to efficiently assist in code writing tasks. 
Moreover, focusing on the keywords of programming-related topics, we find ChatGPT's ability to assist in both developing and analytical work.

In our analysis of topics related to news, we found a notable shift in user interests within our datasets. Prior to April 2023, in our dataset of 52k conversations, users frequently asked about the Russia-Ukraine conflict. However, in the subsequent dataset of 40k conversations, gathered between April and June 2023, the focus shifted to a soccer match between Argentina and Korea, with particular interest in Lionel Messi, the renowned football star. This change in user queries aligns with the timing of these international events, reflecting a clear consistency between global news trends and user inquiries.

\section{Limitations}
The primary limitation of our study is centered around the data source. The data were collected through shareGPT, an optional plug-in for ChatGPT, which not all users choose to install or use to share their conversation records. While we express our sincere gratitude to those who opted to share their records, contributing significantly to our research and enabling the collection of over 92,000 conversation records in six months, it is crucial to recognize the inherent discrepancies between our dataset and the wider ChatGPT user base. The number of users represented in our dataset is considerably smaller than the total number of ChatGPT users. Additionally, commercial constraints prevent us from accessing a more exhaustive dataset for analysis. Nevertheless, this research provides a valuable framework for analyzing ChatGPT user behavior, paving the way for future studies. This groundwork will be particularly beneficial once a more extensive dataset becomes accessible, allowing researchers to conduct more comprehensive research efficiently.

Furthermore, with the diversification of input types accommodated by ChatGPT and analogous software, coupled with an increase in the number of utilized plug-ins, there is an aspiration to develop distinct algorithms for enhanced input type recognition and to conduct more profound research in this domain.

As of the final stages of composing this paper, ChatGPT-4 has introduced image input capabilities, signifying a broader application scope for this tool. We anticipate a more diverse range of user inputs to ChatGPT in the future, aiming to gain deeper insights into evolving user needs and preferences.

\section{Conclusion}
In our study of a newly released dataset featuring ChatGPT user interactions, we conducted a comprehensive analysis of early ChatGPT user interactions to understand the conversation dynamics and user portrait. This included analyzing conversation statistics, sentiment trends analysis, and topic categorization, among other factors.

In our analysis, we discovered that among the 45 languages examined, English-language conversations were predominant, followed by Chinese, Korean, French, and Spanish. Our sentiment analysis revealed that ChatGPT typically produces responses with a positive tone and often steers conversations toward positive or neutral outcomes. Regarding the subjects of these conversations, we noted that they predominantly revolved around topics such as coding, business analysis, news and current affairs, and data analysis.

\bibliographystyle{unsrt}
\bibliography{ref}

\end{document}